\pdfoutput=1
\documentclass[reprint, aps, prd, preprintnumbers, amsmath, amssymb, superscriptaddress, nofootinbib]{revtex4-1}
\usepackage{amssymb,amsmath,bbold,slashed}
\usepackage{graphicx,soul}
\usepackage[usenames,dvipsnames]{xcolor}
\usepackage[unicode=true,pdfusetitle,
 bookmarks=true,bookmarksnumbered=false,bookmarksopen=false,citecolor=Turquoise,
 breaklinks=false,pdfborder={0 0 1},backref=false,colorlinks=true,pdfpagemode=FullScreen]
 {hyperref}

\begin{document}

\preprint{IPMU20-0012}

\title{A Complete Solution to the Strong CP Problem:\\ a SUSY Extension of the Nelson-Barr Model}

\author{Jason Evans}
\email[]{jlevans@sjtu.edu.cn}
\affiliation{T. D. Lee Institute and School of Physics and Astronomy, Shanghai Jiao Tong University, Shanghai 200240, China}

\author{Chengcheng Han}
\email[]{hancheng@kias.re.kr }
\affiliation{School of Physics, KIAS, 85 Hoegiro, Seoul 02455, Republic of Korea}

\author{Tsutomu T. Yanagida}
\email[]{tsutomu.tyanagida@ipmu.jp \\ Hamamatsu Professor}
\affiliation{T. D. Lee Institute and School of Physics and Astronomy, Shanghai Jiao Tong University, Shanghai 200240, China}
\affiliation{Kavli IPMU (WPI), UTIAS, University of Tokyo, Kashiwa, Chiba 277-8583, Japan}
\author{Norimi Yokozaki}
\email[]{yokozaki@tuhep.phys.tohoku.ac.jp }
\affiliation{Department of Physics, Tohoku University, Sendai, Miyagi 980-8578, Japan}

\begin{abstract}
We present a supersymmetric solution to the strong CP problem based on spontaneous CP violation which simultaneously addresses the affects coming from supersymmetry breaking. The generated CP violating phase is communicated to the quark sector by interacting with a heavy quark a la Nelson-Barr. The Majorana mass of the right handed neutrinos is generated by interactions with the CP violating sector and so does not conserve CP. This gives the neutrino sector a non-trivial CP violating phase which can then generate the baryon asymmetry of the universe through leptogeneis. The problematic phase in the supersymmetry breaking parameters are suppressed by appealing to a particular gauge mediation model which naturally suppresses the phases of the tree-level gluino mass. This suppression plus the fact that in gauge mediation all loop generated flavor and CP violation is of the minimal flavor violation variety allows for a complete and consistent solution to the strong CP problem.   
\end{abstract}

\maketitle

\section{Introduction}

Despite the fact that supersymmetry has evaded all observations, it remains a compelling model of nature which ameliorates the naturalness problem, provides a WIMP dark matter candidate, and leads to gauge coupling unification. However, the soft masses problematically contain multiple sources of flavor and CP-violation. Although flavor violation decouples from the standard model (SM) if the SUSY breaking scale is taken to be large, this is not necessarily the case for CP violating effects. This problem is seen in the expression for the low-scale $\bar\theta_{QCD}$,
\begin{eqnarray}
\bar \theta =\theta + {\rm Arg\{Det}(M_u M_d)\}-3  {\rm Arg}(M_{\tilde g})~.
\end{eqnarray}
Clearly,if the mass of the gluino is complex, it will also induce $\bar\theta_{QCD}\sim 1$ independent of the SUSY breaking scale. The situation is further complicated by radiative corrections to $M_u,M_d,M_{\tilde g}$ involving $A$-terms and sfermion mass matrices. The current constraint on $\bar\theta_{QCD}$ comes from the measurement of the neutron EDM giving $\bar\theta_{QCD}<10^{-10}$. This not only constrains the CP-violating phase of the gluino to be less than $10^{-10}$, but it restricts phases in the squark mass matrices to be smaller than $\mathcal O(10^{-8})$. This places rather strong constraints on the mechanism of supersymmetry breaking. This is the essence of the supersymmetric strong CP-problem.

One elegant solution to the strong CP problem is to assume the existence of some global U(1) symmetry~\cite{Peccei:1977hh,Peccei:1977ur}. Under this symmetry, some set of fields charged under SU(3) color transform chirally rendering $\bar\theta$ unphysical. However, it is believed that quantum gravity does not respect global symmetries and this clever solution is sullied by Planck suppressed operators unless they are somehow forbidden up to dimension 10. Although this can be engineered by appealing to discrete gauge symmetries, these models tend to be complicated with little motivation beyond removing the problematic Planck suppressed operators which break the U(1) PQ symmetry. 

A more appealing approach is to assume that CP is an exact symmetry which is spontaneously broken at some higher energy scale. Although this may seem like another problematic global symmetry, it has been shown that the CP symmetry could be a gauge symmetry in extra dimensional models~\cite{Strominger:1985it, Choi:1992xp, Dine:1992ya}. Since gravity respects gauge symmetries, we do not need to worry about the quality of the remaining CP symmetry, an important advantage of these types of models. In these types of models, the phase of the CKM matrix is then generated through interactions of SM quarks with the CP breaking field. An example of how to generate the CKM matrix in these types of models was first shown by Nelson-Barr~\cite{Nelson:1983zb, Barr:1984qx, Barr:1984fh}, their mechanism  in supersymmetry will be discussed below in more details. 

In this work, we will present a consistent model which incorporates the ideas of spontaneous CP violation as a solution to the strong CP problem into supersymmetry in a way consistent with leptogenesis. The paper is organized as follows: we overview the Nelson-Barr mechanism in the framework of supersymmetry present the potential problems of their merging, and then show a complete and consistent model that resolves these problems with some discussion of what the SUSY spectrum looks like.

\section{Supersymmetric NB Models}

The simplest SUSY Nelson-Barr mechanism requires two fields to spontaneously break
CP, $\eta_{1,2}$, with different phases and a pair of down-type chiral multiplets $D$, $\bar{D}$ which mix with the SM d-quarks. The $\eta_{1,2}$, $D$, $\bar{D}$ have charge $-1$ under some $Z_2$ in order to forbid unwanted couplings. The superpotential for this theory takes the form,
\begin{eqnarray}
\hspace{-5pt}
W=  Y_{\alpha, i} \eta_\alpha  D  \bar d_i  + M_D D \bar D + y_{ij} H_d Q_i \bar d_j+ W_{\rm MSSM} \label{eq:NBSup}~,
\end{eqnarray}
where $W_{\rm MSSM}$ is the MSSM superpotential. After CP is broken, the mass matrix of down-type quarks become
 \begin{eqnarray}
 \mathcal{M} =\left(
   \begin{array}{cc}
    m_d & B\\
    0     & M_D \\
   \end{array}
 \right),~  m_d \equiv y v_d; B_i= Y_{\alpha, i} \eta_\alpha　~.
 \end{eqnarray}
 Since the only source of tree-level CP violation in $\mathcal{M}$ is $B_i$,  Arg$[ {\rm det} \mathcal M]=0$ and so $\bar \theta=0$ is maintained. If $M_D \lesssim B_i$, a large CKM phase is generated when the $D$ and $\bar{D}$ are integrated out. In the supersymmetric limit, $\bar \theta$ is protected by the non-renormalization theorem. Thus, only supersymmetry breaking effects can spoil this solution.

\begin{figure}[ht]
\centering
\includegraphics[width=3.5in]{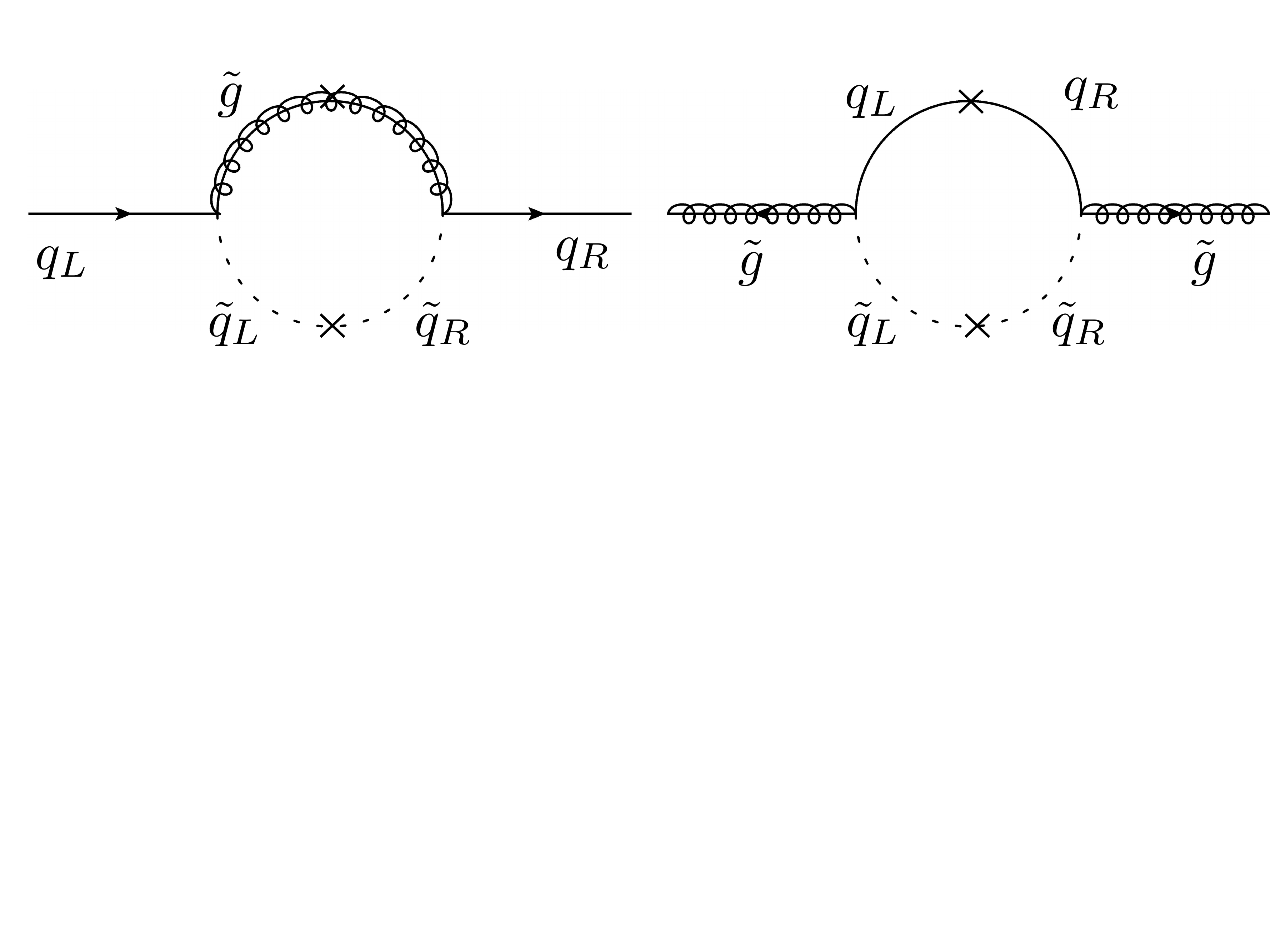}
\vspace{-4.3cm}
\caption{\label{fig:FeyDia}Loop diagrams contribution to $\bar \theta$.}
\label{plot1}
\end{figure} 
 
This simple story is, however, complicated by supersymmetry breaking. In models based on supergravity, there are two sources of supersymmetry breaking which can contribute to the gluino mass, the F-term of some field and the gravitino mass. If CP is broken above the scale where supersymmetry breaks, both sources of SUSY breaking will, in general, be complex. In gravity mediated models, the F-term is usually the dominate contribution to the gluino mass and so must be real. However, even the subdominant piece coming from $m_{3/2}$, due to anomaly mediation, is problematic since it will only be loop suppressed relative to the dominate piece. If $m_{3/2}$ has a generic phase, it will lead to $\bar\theta\sim 10^{-2}$. Thus, for gravity mediated models to work both $m_{3/2}$ and the F-term that breaks supersymmetry must be real to a very high level of precision. 

Models based on pure anomaly mediation do better than gravity mediated models, since the F-term contribution to the gluino masses is forbidden by some symmetry.  In this case, the gluino mass is proportional to the gravitino mass, which must still be real. This is a much more manageable feat. However, pure anomaly mediation has a tachyonic slepton problem\footnote{This problem may be solved by allowing the supersymmetry breaking field to interact with the Higgs superfields in the K\"alher~\cite{Yin:2016shg}.}. The most obvious and natural solution to this problem is to relax the ad hoc assumption about the K\"ahler potential
With a more generic K\"ahler, the sfermion masses can be generated from both the F-term of the SUSY breaking and the gravitino mass. In this case, the sleptons are no longer tachyonic and have masses that are much larger than the gluino masss. If the F-term breaking SUSY is complex, this will leads to large phases in the sfermion masses matrices. One might hope that since the sfermions are much heavier than the gluino, these phases would decouple and have little affects on $\bar\theta$. As we will see below, in Nelson-Barr models, decoupling the sfermions does not remove the effects of these phases.

The problematic contribution to $\bar\theta$ in models like pure gravity mediation come from the loop corrections to gluino mass and down quark mass. The dangerous supersymmetry breaking terms are
\begin{eqnarray}
V_{soft}&\supset & \bar d_i^\dagger m_{\tilde d_{ij}}^2 \bar d_j+Q_i^Tm_{Q_{ij}}^2Q^*_j+m_{\tilde D}^2|D|^2+m_{\tilde{\bar D}}^2|\bar D|^2\nonumber \\
&&\quad \ \ + A_{d_{ij}}H_d Q_i\bar d_j+B_q\mu q\bar q+A_{Y_{\alpha i}}H_d \eta_\alpha \bar d_j ~.
\end{eqnarray}

Assuming a real tree level gluino mass, the diagram for the dangerous one loop correction to gluino mass is found in Fig.(\ref{fig:FeyDia}) and leads to the following important constraints 
\begin{eqnarray}
\frac{\alpha_s}{4\pi}\frac{{\bf Im}(B_i^\dagger  A_{Y_{\alpha i}}\eta_\alpha)}{(|B_j|^2+M_D^2){\rm Re}(m_{\tilde g})} \lesssim 10^{-8},\label{eq:higcon}
\end{eqnarray}
where $m_{\tilde g}$ is the gluino mass. This constraints require that the A-terms be universal or adhere to some kind of minimal flavor violation. In models based on gravity mediated supersymmetry breaking, this will not be the case when all Planck suppressed higher dimensional operators are included in the K\"alher. Thus, even if the gluino mass is real at tree level, further constraints are needed to solve the strong CP problem.  In pure gravity mediation, the constraint in Eq. (\ref{eq:higcon}) on $\bar\theta$ does not get better, since this constraints does not change as the scale of the soft masses is changed

The other problematic radiative corrections come from loop corrections to the quark mass matrices . The Feynman diagram is shown in Fig. (\ref{fig:FeyDia}) . These corrections to the quark mass matrix is even more dangerous since they only depends on the diagonal part of the A-terms
This contribution gives a constraint 

\begin{eqnarray}
\hspace{-5pt}
\frac{\alpha_s}{4\pi}\frac{A_d{\bf Im}(B^\dagger m_d\delta m_Q^2m_d^{-1} \left(\delta m_{\tilde d}^2-\delta m_{\tilde D}^2\right)B} {m_{SUSY}^5(|B_j|^2+M_D^2)}\lesssim 10^{-11}
\label{eq:massplit},
\end{eqnarray}
where $A_d$ is the diagonal part of the A-terms and $\delta m_{\tilde d}^2$, $\delta m_{\tilde D}^2$ are the variation of these masses from their universal value, and $m_{SUSY}$ is the scale of the sfermion masses. Again, this constraint is not made better by decoupling the soft masses. Thus, even models like pure gravity mediation have problematic radiative corrections to $\bar\theta$.

Gauge mediation models fare better~\cite{Hiller:2002um, Dine:1993qm}. In gauge mediation models, the SUSY breaking scale is much lower, scaling with the mediation scale. This reduction in the SUSY breaking scale also suppresses the gravitino mass since they are related through the cosmological constant. This helps the strong CP problem in three ways. First, the reduction in the SUSY breaking scale neutralize the effects of the Planck suppressed operators, since their contribution will be suppressed by a factor of $(M_*/M_P)^{(n-4)}$ where $n$ is an integer corresponding to the dimensionality of the operator. Second, the reduction in the gravitino mass makes the anomaly mediated contribution benign, if the mediation scale is low enough. Third, since the A-terms are loop suppressed relative to the other soft masses and are proportional to the MSSM Yuakwa couplings, the radiative corrections to $\bar\theta$ are also suppressed.

Although the dominant source of the soft masses is flavor universal in gauge mediation, the NB sector breaks this universality. The contribution comes from inserting an $\eta_\alpha$ and $D$ is loop into the two-loop gauge mediated soft mass diagram.  This generates a flavor non-universal contribution to the soft masses proportional to $y_{D_{\alpha, i}} y^\dagger_{D_{\alpha, j}}$. For $y_{D}\sim 1$, this leads to the constraint
$B_i\sim M_D \lesssim 10^{2} M_*$, where $M_*$ is the messenger scale.
If this conditions is satisfied, the constraint in Eq. (\ref{eq:massplit}) will also be satisfied.

\begin{table*}[t]
\caption{Charge assignments for all the fields. \label{tb:charassign} }
\begin{center}
\begin{tabular}{|c|c|c|c|c|c|c|c|c||c||c|c|c|c|c|c|c|}
\hline  ~&  \multicolumn{9}{|c||}{Matter fields}            &  R-breaking sector  &  \multicolumn{3}{|c|}{ SUSY breaking sectors }     \\    
\hline  Fields &  ~~${\bf \bar{5}}$~~ & ~~{\bf 10}~~&~~ $N^c$~~ & ~~$H_u$~~ &~~ $ H_d$~~&~~$5^\prime$  ~~&~~ $\bar{5}^\prime$ ~~&~~ $\eta_{1,2}$ ~~&~~ $Y_{1,2}$~~ &~~ $S$~~ &~~ $S^\prime$~~ 
&~~ $S^{\prime \prime}$~~ &$\Psi_{1,2}$/$\bar\Psi_{1,2}$ \\
\hline $Z_{4}$ & $i$ & $i$ & $i$ & $-$ & $-$ & $i$ & -$i$   & $-$ & $+$  & $+$   & $+$  & $+$     & $+$          \\
\hline R-charge &    0    & 0  &0 &2 &2  & 0 & 0 & 2  &2 &2  & 2 & 2  & 0   \\
\hline $Z^\prime_3$ &   1 &  2  & 0  & 2  & 0 &  2  & 1  & 0   &  0  &0   &  0 & 1   &0   \\
\hline $Z_3$ &     \multicolumn{9}{|c||}{0}   & 0 &1   & 0   &1   \\
\hline 
\end{tabular}
\end{center}
\label{table1}
\end{table*}%


Even though gauge mediation solves many of the problems associated with the supersymmetric strong CP problem, the model is far from complete. For example, the baryon asymmetry must be generated and requires an additional source of CP violation beyond that in the quark Yukawa couplings.  The simplest way to generate the baryon asymmetry of the universe (BAU) is through leptogenesis~\cite{Fukugita:1986hr}. This requires adding heavy right-handed neutrinos, $N^c_i(i=1,2,3)$,  that have CP violating interactions. These heavy right handed neutrinos can also explain the smallness of the neutrino masses via the seesaw mechanism~
\cite{Minkowski:1977sc, Yanagida:1979as, Glashow:1979nm, Ramond:1979py, GellMann:1980vs}. Since NB theories require CP be violated spontaneously, CP violation must be communicated to the neutrino sector as well. Because the right-handed neutrinos are singlets, it may seem trivial to generate CP violation in the neutrino sector by directly coupling the CP breaking fields,$\eta_\alpha$, to the right handed-neutrinos through the interaction $Y^{ij}_\alpha\eta_\alpha N^c_iN^c_j$. However, this operator is problematic, since it violates the $Z_2$ symmetry protecting the NB theory from other operators spoiling the mechanism. Fortunately, as we will see below, other discrete symmetries can be used to forbid the problematic operators from the NB theory and still allow for this CP violating interaction in the lepton sector.

Additional complications arise from Planck suppressed operators, which are not fixed by the global symmetries. The operators $\frac{1}{M_P} \eta_\alpha \eta_\beta D \bar{D} $ and  $\frac{1}{M_P}\eta_\alpha H_d {Q}_i \bar{D}$, for example, would introduce complex contributions to $m_d$ and $M_D$ respectively.  If either of these masses has a large complex component, it would lead to Arg$[ {\rm det} \mathcal M]\ne 0$ and reintroduce the strong CP problem. These operators are suppressed by the Planck scale and so can be sufficiently suppressed. However, they require $\eta_\alpha\lesssim 10^8$ GeV~\cite{Dine:2015jga}. Combining this with the constraint on the messenger scale discussed above, we find that $M_*\lesssim 1$ PeV. Because we want the soft masses of the MSSM to be of order a few TeV, this bound on the messenger scale translates into a bound on the gravitino mass of $m_{3/2}\sim M_*^2/M_P\lesssim 1$ MeV. With a gravitino mass this small, the gravity mediated contribution to the the soft masses will not significantly affect the $\bar\theta$. The messenger scale, $M_*$, is also bounded from below by experimental constraints on soft masses and Higgs boson mass.
This places a lower bound on the gravitino mass of order few eV~\cite{Ibe:2016kyg,Hook:2015tra,Hook:2018sai}.

Another difficulty of the NB theory is the breaking of CP tends to generate a domain wall problem.  Since the superpotential preserves CP, the potential is invariant under $\langle \eta_\alpha \rangle  \rightarrow \langle \eta_\alpha \rangle^*$.  The domain wall problem can be avoided if the reheat temperature is low enough that CP is not restored after inflation, i.e.  $T_{rh} < \langle \eta_\alpha \rangle \lesssim 10^8$ GeV. However, thermal leptogenesis generally requires $T_{rh} \gtrsim 10^9$ GeV ~\cite{Giudice:2003jh}. In the model we discuss below, the problematic Planck suppressed operators are forbidden and the upper bound on $\langle \eta_\alpha \rangle$ is removed, allowing for a viable thermal leptogenesis model.  


\section{A Viable Supersymmetric Nelson-Barr Model}
In the last section we highlighted the difficulties of combining SUSY and the NB mechanism. In this section, we will present a model that can alleviate all of these problems and show that the NB mechanism can be consistent with a complete supersymmetric model of nature.

First of all,  let
us describe the model. The matter content and the charge assignments are listed in Table (\ref{tb:charassign}). The superpotential for the Nelson-Barr theory consistent with these symmetries is given by 
\begin{eqnarray}
W&=& y_d H_d {\bf \bar{5}}_i  {\bf 10}_j  + y_u H_u {\bf10 }_i {\bf10 } _j+  y_\nu H_u N_i^c {\bf \bar{5} }_j    
\nonumber \\
&&\quad  + \lambda_D S 5^\prime \bar{5}^\prime + y_D \eta_\alpha 5^\prime {\bf \bar{5}}_i + y_N \eta_\alpha N^c_i N^c_j \label{eq:NBMSSMSup}~,
\end{eqnarray}
where $10\ni \{
Q,~\overline{U},~V_{CKM}\overline{E}
\}$, $\overline{5}\ni\{
\overline{D},~L
\}~$, $N_i$ are the right-handed neutrinos, $5',\bar 5'$ contains the NB quark superfield, $S$ is a singlet under the SM gauge symmetries and is necessary to  breaks the $Z_{4R}$ allowing a mass for $5',\bar 5'$, and $\eta_\alpha$ is the field responsible for breaking CP\footnote{We use SU(5) notation for presentational simplicity. We do not consider a full SU(5) theory.}. 

Our superpotential is in Eq. (\ref{eq:NBMSSMSup}) and has been modified from the simple NB model above.  First, we have introduced right-handed neutrinos and coupled them to the CP breaking fields
, $\eta_\alpha$. This can be done in our model because the $Z_2$ symmetry has been replaced by an anomaly free $Z_4$
. The $Z_{4R}$ symmetry in Table (\ref{tb:charassign}) is only anomaly free if we extend the NB quark superfields to a $5'$ and $\bar 5'$\textcolor{red}{~\cite{Kurosawa:2001iq}}. This inclusion of a vector pair of lepton doublets, $L'$ and $\bar L'$, leads to additional CP violation in the lepton sector a la the NB mechanism. Another important feature of this $Z_{4R}$ is that it forbids the problematic operators $\frac{1}{M_P} \eta_\alpha \eta_\beta D \bar{D} $ and  $\frac{1}{M_P}\eta_\alpha H_d {Q}_i \bar{D}$. With these operators forbidden, the scale of CP breaking can be pushed up and in turn the reheat temperature. A higher reheat temperature then salvages leptogenesis. Thus, all necessary CP violating phases can be generated in this model.

One challenge coming from this $Z_{4R}$ symmetry is the that it forbids a supersymmetric mass term for both the $H_uH_d$ and $5^\prime {\bar 5}^\prime$. The $H_uH_d$ mass term will discuss later. For the $5^\prime, {\bar 5}^\prime$, this problem can be over come by introducing a new singlet field which breaks R-symmetry. This is the field $S$ in Eq. (\ref{eq:NBMSSMSup}). The $R$-symmetry is broken supersymmetrically with a real vacuum expectation value when $S$ gets a vev. The superpotential which spontanesouly breaks the $Z_{4R}$ symmetry is
\begin{eqnarray}
W_{S} &=&  \mu^2 S - \frac{1}{3}\lambda S^3 ~.
\end{eqnarray}
Because the $\langle W \rangle \ne 0$ when the potential is minimized, this will generate a gravitino mass and plays a role in cancelling the cosmological constant\footnote{Since this superpotential has two degrees of freedom, $\mu$ and $\lambda$, one can be used to cancel the cosmological constant.}. 

Problematically, the interaction $S \eta_\alpha \eta_\beta$ is allowed by all the symmetries in Table(\ref{tb:charassign}). If present, this operator would results in mixing between $S$ and $\eta_\alpha$  and the phase of $M_D$ would be too large. However, this operator can be suppress if $S$ and the other field live on separate branes with $5^\prime, {\bar 5}^\prime $ living in the bulk. This allows $S$ to give $5^\prime, {\bar 5}^\prime $ a supersymmetric mass and suppresses $S \eta_\alpha \eta_\beta$.

Next, we discuss the spontaneous breaking of the CP symmetry. The superpotential for this breaking is 
\begin{eqnarray}
W_{\eta} &=& \lambda_1Y_1 (\eta_1^2 + M_1^2)  +  \lambda_2Y_2 ( \eta_2^2 - M_2^2)  ~,
\end{eqnarray}
where $Y_i$ are singlets introduced to assist in breaking the CP symmetry. We have assumed that some of the couplings between $Y_i$ and $\eta_\alpha$ are small and can be ignored to show that CP can indeed be spontaneously broken from this superpotential, with the $\eta_\alpha$'s having different vevs. However, even if we allow these couplings to be large, it is not difficult to find a combination of couplings which breaks CP with $\eta_1$ and $\eta_2$ having different phases.

As we discussed above, the soft masses must be quite diagonal and real to avoid generating a large $\bar\theta$. If the F-term which couples to the messengers of gauge mediation is real, all flavor and CP violation in the soft masses will be proportional to the Yukawa couplings. To ensure this is the case, we use the Evans-Sudano-Yanagida model~\cite{Evans:2010ru}. In this model, a singlet field charged under some $Z_3$,  $S^\prime$, interacts with two pairs of  messengers $ \Psi_i, \bar \Psi_i$, with the matter content of $5, \bar 5$ representation respectively. The allowed interactions are then
\begin{eqnarray}
W_{S^\prime} &=& \frac{\lambda^\prime}{3} {S^\prime}^3 + \kappa S^\prime \Psi_i \bar \Psi_i~,
\end{eqnarray}
where $\lambda'$ is taken to be negative by a field redefinition of $S'$.
In order for this mechanism to work, $S'$ must also have a tachyonic soft masses\footnote{For a way to dynamically generate this tachyonic mass, see the appendix of ~\cite{Evans:2010ru}.}
\begin{eqnarray}
V_{soft}\supset -m_S^2|S^\prime|^2 ~,
\end{eqnarray}
where we have written the soft mass so that $m_S^2$ is positive. This mechanism also requires corrections to the K\"ahler of the form
\begin{eqnarray}
\Delta K= -\frac{h}{\Lambda^2}|S^\prime|^4 ~,
\end{eqnarray}
where $h$ is a dimensionless coupling and is assumed to be positive, and $\Lambda$ is some scale lower than the Planck scale.  The resulting potential, to leading order in SUSY breaking, is
\begin{eqnarray}
\hspace{-15pt}
V\supset  -m_S^2|S^\prime|^2 + |\lambda' S^\prime|^2+ 4\lambda' m_{3/2}\frac{h}{\Lambda^2}|S^\prime|^5\cos3 \delta_{S^\prime} ~,
\end{eqnarray}
where $\delta_{S^\prime}$ is the phase of ${S^\prime}$. As is clear from this potential, there are three minimum $\delta_{S^\prime}=0,\frac{2}{3}\pi,\frac{4}{3}\pi$.  Since the gaugino masses for this model are 
\begin{eqnarray}
m_i=-\frac{\alpha_i}{4\pi}\lambda'|{S^\prime}|e^{-i3\delta_{S^\prime}}~,
\end{eqnarray}
they will be real for all three minima.  However, these three degenerate minima, due to the $Z_3$, have introduced a domain wall problem. Fortunately, the $Z_3$ is broken by gravity lifting the degeneracy of the vacua, with $\delta_{S^\prime}=0$ remaining as the true vacuum of the theory\footnote{Which minima is the true vacuum depends on the sign of the Planck suppressed operator breaking the $Z_3$, $\Delta W=\kappa S'^4/M_P^2$\cite{Vilenkin:1981zs}. In order to solve the domain wall problem, we must appropriately choose the signs of $\kappa$, $\lambda'$, and $h$ so that $\delta_{S'}=0$ is the true minimum.} . 

Here we discuss how to generate the Higgs bilinear mass. As was done in~\cite{Evans:2010ru}, an additional field $S''$ can be added which has a SUSY scale vev and F-term which are generated in a similar manner as $S^\prime$. The vacuum structure of this theory is such that the phases of $B$ and $m_{\tilde g}$ are correlated and so both are real. This mechanism solves the $B,\mu$ problem while suppressing any phases which would reintroduce the strong CP problem.

Lastly, we mention the parameter space of this model, which is consistent with all observations. Since it is a fairly generic gauge mediation model, it requires a quite heavy SUSY spectrum to get the Higgs mass heavy enough. However, since the error bars on the calculation of the Higgs mass are quite large, it is possible that the HL-LHC could be sensitive to the heavy Higgs of this model. 


\section{Conclusions}
In this letter we have examined the Nelson-Barr solution to the strong CP problem in the context of supersymmetry. By generalizing the global symmetries of past supersymmetry Nelson-Barr models and including gauge R-symmetries, the problematic operators, which either tightly constrain or forbade past models, can be alleviated. Furthermore, this generalization allows for a Nelson-Barr like mechanism to generate a CP violating phase in the lepton sector which could explain the baryon asymmetry of the universe. It also requires the right-handed neutrino masses be generated by the field that spontaneously breaks CP. This not only gives another source of CP violation to assist in leptongenesis, but it also correlates these scales in a unique way. To complete our model, we have shown that if nature relies on a particular form of gauge mediation to generate the MSSM soft masses, no additional flavor or CP violation is generated in the low-scale. Thus, this is a consistent model of SUSY which can explain the baryon asymmetry of the universe, the strong CP problem, and has a completely consistent LHC spectrum.




\begin{acknowledgments}
C.C.H, N.Y., and T.T.Y would like to thanks Peter Cox for useful discussion during the early stages of this work. J.L.E and T.T.Y. would like to thank the people of IPMU for their hospitality, which facilitated the completion of this work. This work is supported by Grants-in-Aid for Scientific Research from the Ministry of Education, Culture, Sports, Science, and Technology (MEXT), Japan, No.~26104001 (T.T.Y.), No.~26104009 (T.T.Y.), No.~16H02176 (T.T.Y.), No.~17H02878 (T.T.Y.), 
No.~15H05889(N.Y.), No.~15K21733(N.Y.), and No.~17H02875 (N.Y.),
and by the World Premier International Research Center Initiative (WPI), MEXT, Japan (T.T.Y.). 
\end{acknowledgments}

\end{document}